**Water-enhancing gels exhibiting heat-activated formation of silica aerogels for protection of critical infrastructure during catastrophic wildfire**

*Changxin Dong, Andrea I. d'Aquino, Samya Sen, Ian A. Hall, Anthony C. Yu, Jesse D. Acosta, and Eric A. Appel\**

C. Dong, A. I. d'Aquino, S. Sen, A. C. Yu, Prof. E. A. Appel
Department of Materials Science & Engineering, Stanford University, Stanford, CA 94305, USA
E-mail: eappel@stanford.edu

I. A. Hall, Prof. E. A. Appel
Department of Bioengineering, Stanford University, Stanford, CA 94305, USA

J. D. Acosta
Department of Natural Resource Management & Environmental Sciences, California Polytechnic State University, San Luis Obispo, CA 93407, USA

Prof. E. A. Appel
Stanford ChEM-H Institute, Stanford University, Stanford, CA 94305, USA

Prof. E. A. Appel
Woods Institute for the Environment, Stanford University, Stanford, CA 94305, USA

Prof. E. A. Appel
Department of Pediatrics - Endocrinology, Stanford University School of Medicine, Stanford, CA 94305, USA

**Abstract**

A promising strategy to address the pressing challenges with wildfire, particularly in the wildland-urban interface (WUI), involves developing new approaches for preventing and controlling wildfire within wildlands. Among sprayable fire-retardant materials, water-enhancing gels have emerged as exceptionally effective for protecting civil infrastructure. They possess favorable wetting and viscoelastic properties that reduce the likelihood of ignition, maintaining strong adherence to a wide array of surfaces after application. Although current water-enhancing hydrogels effectively maintain surface wetness by creating a barricade, they rapidly desiccate and lose efficacy under high heat and wind typical of wildfire conditions. To address this limitation, we developed unique biomimetic hydrogel materials from sustainable cellulosic polymers crosslinked by colloidal silica particles that exhibit ideal viscoelastic properties and facile manufacturing. Under heat activation, the hydrogel transitions into a highly porous and thermally insulative silica aerogel coating in situ, providing a robust protective layer against ignition of substrates, even when the hydrogel fire suppressant becomes completely desiccated. By confirming the mechanical properties, substrate adherence, and enhanced substrate protection against fire, these heat-activatable biomimetic hydrogels emerge as promising candidates for next-generation water-enhancing fire suppressants. These advancements have the potential to dramatically improve our ability to protect homes and critical infrastructure during wildfire.

# 1. Introduction

Catastrophic wildfires have grown progressively more frequent and severe in the United States, Europe, and Australia due to a combination of factors such as climate change, which has led to hotter and drier seasons, historical fire suppression policies that have allowed fuels to accumulate over nearly a century, and inadequate vegetation management to remove these accumulated fuels from the landscape. Wildfire incidents result in devastating losses to the economy, critical infrastructure, wildland resources, and the lives and livelihoods of people living within the wildland-urban interface (WUI).[1] To effectively address and combat wildfires, new classes of environmentally friendly wildland fire retardants must be developed that can improve wildfire management efforts.

Wildland fire chemical systems are classified by the US Forest Service (part of the US Department of Agriculture; USDA) as either (i) long-term retardants, (ii) foam suppressants, or (iii) water-enhancing gels.[2] Long-term retardants are supplied as liquid or solid concentrates and are typically dropped by aircraft over target landscapes after full constitution through dilution with water. The retarding effects of these products are produced by fire-retarding chemicals such as ammonium phosphates, which serve to decrease the intensity of burning of treated vegetation, and these retardants maintain their efficacy as long as the chemical residue remains on vegetation.[3] Foams and water-enhancing gels both act as short-term fire suppressants only, meaning the fire resistance is lost once the water has evaporated from the applied materials. Many foams have historically contained perfluorinated surfactant agents that enhance foam adherence and retention to surfaces, as well as water penetration into woods.[4] Water-enhancing gels typically comprise superabsorbent polymers that strongly retain water and improve surface wettability, thus functioning effectively as fire suppressants and barricades to protect critical infrastructure in the wildland and structures such as homes in the WUI.[5] Unfortunately, while perfluorinated surfactants are highly efficacious first suppressants, they exhibit significant bio-accumulation and cause severe environmental toxicities. Thus, water-enhancing gels constitute a more environmentally friendly wildland fire suppression strategy as their properties can be tuned to exhibit excellent substrate adherence and water retention.[6-9]

Many biocompatible hydrogel materials have been developed and exploited in a variety of applications, including drug delivery,[10-12] water treatment,[13-15] and medical implants.[16-18] Specifically relevant for fire suppression efforts, the cross-linked polymer structure of hydrogels enables them to effectively absorb and retain substantial amounts of water. The exceptional water storage capabilities of hydrogels enable extended cooling and wetting duration when applied to vegetation or other substrates. Yet, while hydrogels are an appealing alternative to conventional long-term retardants and foam suppressants on account of their ability to cool and insulate to protect substrates against fires,[19] these materials are limited to a very short application timeframe during typical wildfire weather. Once these materials dry under high heat and wind conditions, they are no longer effective, which severely limits their practical usage. Water-enhancing gel technologies, therefore, require critical advancements to retain their effectiveness during wildfire.

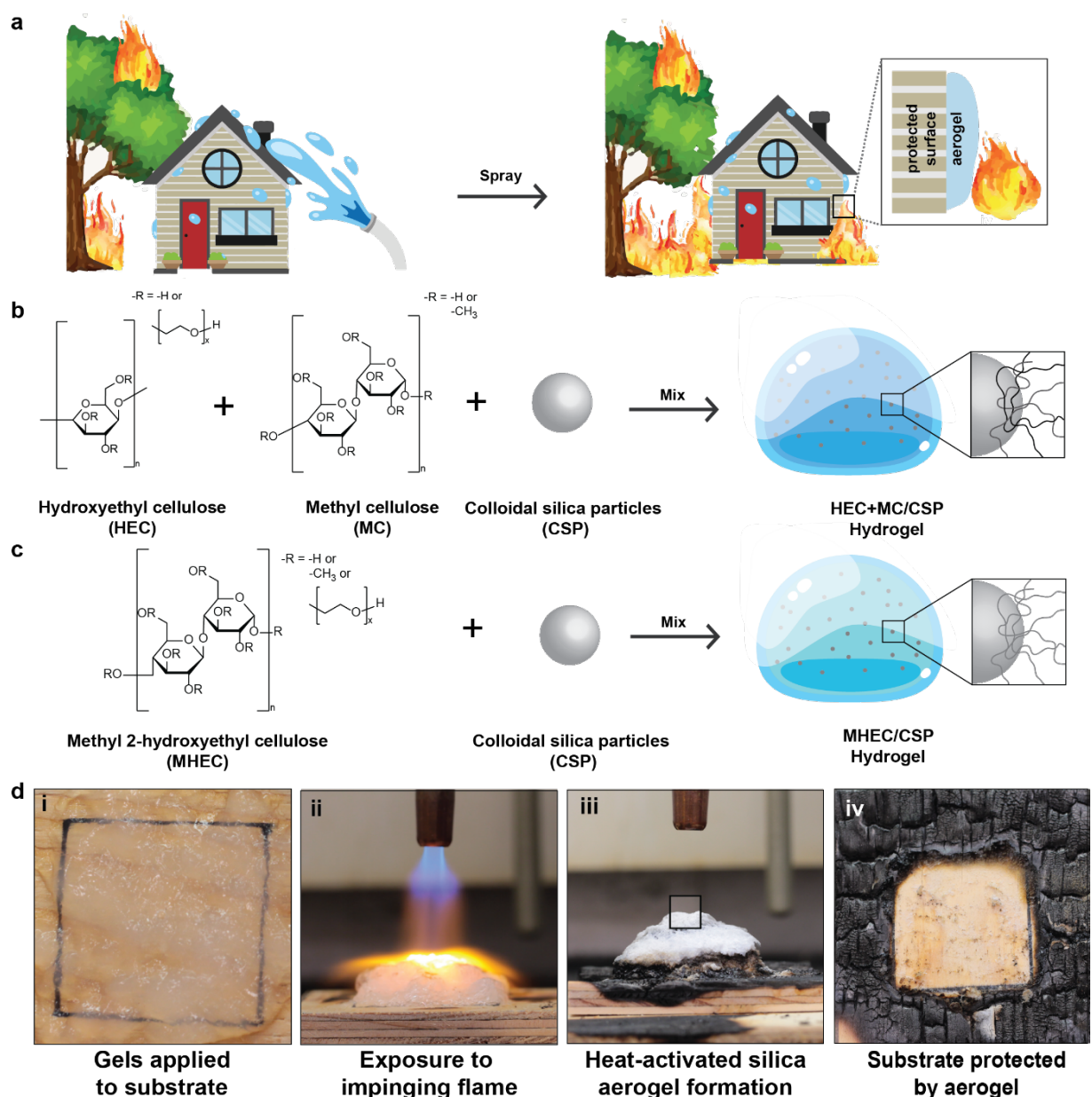

**Figure 1. Schematic of PP hydrogel functions as sprayable fire-retardant gel. a.** Illustration of water-enhanced fire gel effectively repelling wildfire from a house at WUI after prophylactic spray treatment, **b.** methyl-2-hydroxyethyl cellulose (MHEC) adheres to colloidal silica particles (CSP) by interfacial adsorption, **c.** hydroxyethyl cellulose (HEC) and methylcellulose (MC) adhere to CSP in a multivalent, non-covalent manner to form water-enhancing gel, and **d.** Images of (i) applying fire gel onto wood substrate, (ii) burning the gel with a fire torch, (iii) forming silica aerogel, (iv) revealing the intact, uncharred wood under the silica aerogel protective layer.

In this work, we introduce a water-enhancing hydrogel platform that transforms during heat activation into a silica aerogel with exceptional insulating properties, extending the effective window of substrate protection from ignition. These materials exhibit barricade properties improved by over 10-fold compared to existing commercial water-enhancing gel products. These hydrogels are formed by polymer-particle (PP) interactions between cellulosic biopolymers and colloidal silica particles. We show that under flame impingement, water within these materials is rapidly lost, leading to the formation of a robust solid silica network that acts as a robust physical barrier against fire. Our experiments reveal that the PP interactions between the cellulose chains and silica particles play a crucial role in effectively forming the heat-insulative aerogel shield upon water evaporation. This unique mechanism to prolong protective effectiveness is highly distinguishing amongst wildfire suppressant technologies, effectively addressing a crucial gap in our ability to combat wildfires and protect critical infrastructure.

## 2. Results and Discussion
### 2.1. Polymer-particle hydrogel synthesis

Our group previously reported the development of a dynamically crosslinked, biomimetic viscoelastic fluid platform for the encapsulation and sustained retention of ammonium poly(phosphate) based wildland fire retardants on vegetation of interest.[20] These viscoelastic fluid materials leverage physical crosslinking interactions between cellulosic biopolymers and colloidal silica particles, which we call Polymer-Particle (PP) interactions. These materials are readily manufactured at scale, making them amenable to diverse applications from pipeline maintenance, recovery of food and beverage products, and as carrier fluids for wildland fire retardants.[21, 22] Indeed, these materials form the basis for the commercial phosphate-based wildfire prevention product called Phos-Chek® FORTIFY® (Perimeter Solutions).

As mentioned above, in scenarios where wildfires approach the Wildland-Urban Interface (WUI), water-enhancing gels that don't carry retardant chemicals (e.g., ammonium phosphates) are often sprayed onto structures to provide preventive and fire-suppressing effects (shown schematically in **Figure 1a**). In this work we sought to develop PP hydrogel materials for use as water-enhancing gels for protection of structures during wildfire. The unique properties of these materials arise from their biomimetic, dynamic, and multivalent interactions that are self-assembled through selective absorption of cellulose chains onto silica particles (**Figures 1b** and **1c**), generating moldable polymer networks exhibiting tunable dynamic mechanical properties and flow properties that can meet various engineering requirements for processes such as injection, pumping, or spraying. We envisioned that hydrogel formulations could be developed that adhere effectively to substrates and which would experience water loss under heat exposure, leaving a silica-based aerogel-like structure that would highly insulative owing to its refractory properties, thereby protecting the underlying substrate (**Figure 1d**).

As PP hydrogels rely on self-assembly of polymer-particle interactions between cellulosic biopolymers and colloidal silica particles (CSP), they are synthesized in a facile manner by simply mixing aqueous solutions of the cellulosic polymers with aqueous dispersions of CSPs (Ludox® TM-50). CSPs form stable dispersions above pH = 9 as there is sufficient density of silanolate moieties present on the particle surface leading to electrostatic repulsion that ensures suspension stability.[23] Upon mixing at appropriate solids content and stoichiometry, the cellulose chains rapidly adhere to the particles,[24] leading to the formation of a robust, dynamically crosslinked hydrogel. The transient nature of the polymer adsorption to the particles imbues the networks with desirable physical behaviors including shear-thinning and rapid self-healing that are useful for facile processing and tunable substrate adherence.[20] The rheological characteristics of these materials are highly tunable based on both the solids content and stoichiometry of the polymer-particle formulation. Notably, cellulose derivatives and colloidal silica are low-cost, environmentally friendly, and biocompatible,[25, 26] making these materials readily suitable for large-scale production.[21]

In this study, we evaluated two PP hydrogel formulations comprising CSPs and either a mixture of hydroxyethylcellulose (HEC) and methylcellulose (MC) (HEC+MC/CSP; **Figure 1b**), or methyl 2-hydroxyethyl cellulose (MHEC) (MHEC/CSP; **Figure 1c**). These formulations are denoted throughout as X-Y, where X refers to the total weight percent loading of polymer and Y refers to the weight percent loading of CSPs. These cellulosic derivatives widely used as thickening agents and in applications for emulsification, bubble formation, water retention, and stabilization in the pharmaceutical, cosmetic, and construction industries.[27-29] Here, we characterized the rheological properties of PP hydrogel formulations comprising these different materials materials, examined the *in situ* formation of silica aerogels upon heat exposure, and evaluated their effectiveness in fire retardation (*vide infra*).

## 2.2. Rheological Characterization of PP and AquaGel-K Gels

A commonly used commercial water-enhancing gel is Phos-Chek® AquaGel-K, consisting primarily of superabsorbent poly(propenoic acid) polymers.[30] AquaGel-K is supplied as a dry powder and instantly forms a gel when mixed with water at a concentration of 0.3-0.5 wt.%. We first compared the rheological properties of various PP hydrogel formulations investigated in this study to those of commercial AquaGel-K gels (**Figure 2**). To quantify the linear viscoelastic properties of the various hydrogel formulations evaluated in this study, as well as analyze the effect of additives, we conducted frequency sweeps in small amplitude oscillatory shear (SAOS) tests (Figure 2).[31, 32] We used values of $G'$ (the shear storage modulus) to describe the relative stiffness of the gels, while we used values of tan($\delta$) ($G''/G'$) calculated from SAOS data at a frequency of 1 rad/s as a metric of the relative solid-fluid nature of the materials (**Figure 2a, b**), where higher values of tan($\delta$) correspond to more dynamic and fluid-like behaviors.[33] AquaGel-K (0.5%) exhibited consistently higher values for both storage ($G'$) and loss ($G''$) moduli at all frequencies ($G'$ = 456.8 Pa, tan($\delta$) = 0.168), indicating higher overall stiffness and greater elasticity (i.e., lower tan($\delta$)) respectively compared to HEC+MC/CSP 1-5 ($G'$ = 46.4 Pa, tan($\delta$) = 0.219) and MHEC/CSP 1-5 ($G'$ = 25.6 Pa, tan($\delta$) = 0.400) hydrogel formulations (**Figure 2a, b**). We see that both PP hydrogel formulations exhibited higher tan($\delta$) compared to AquaGel-K, suggesting that the PP formulations absorb and dissipates deformation energy more efficiently, thus making them more "flowable" or fluid-like.

To quantify the nonlinear flow properties of the gels, we collected steady shear flow curve data for all hydrogel formulations (**Figure 2c, d**). While all the PP hydrogels exhibited slightly lower stiffnesses than AquaGel-K in SAOS experiments, these materials display similar shear-dependent viscosity curves (**Figure 2c**). This observation suggests these materials will have very similar flow behavior during application, which is typically by spraying.

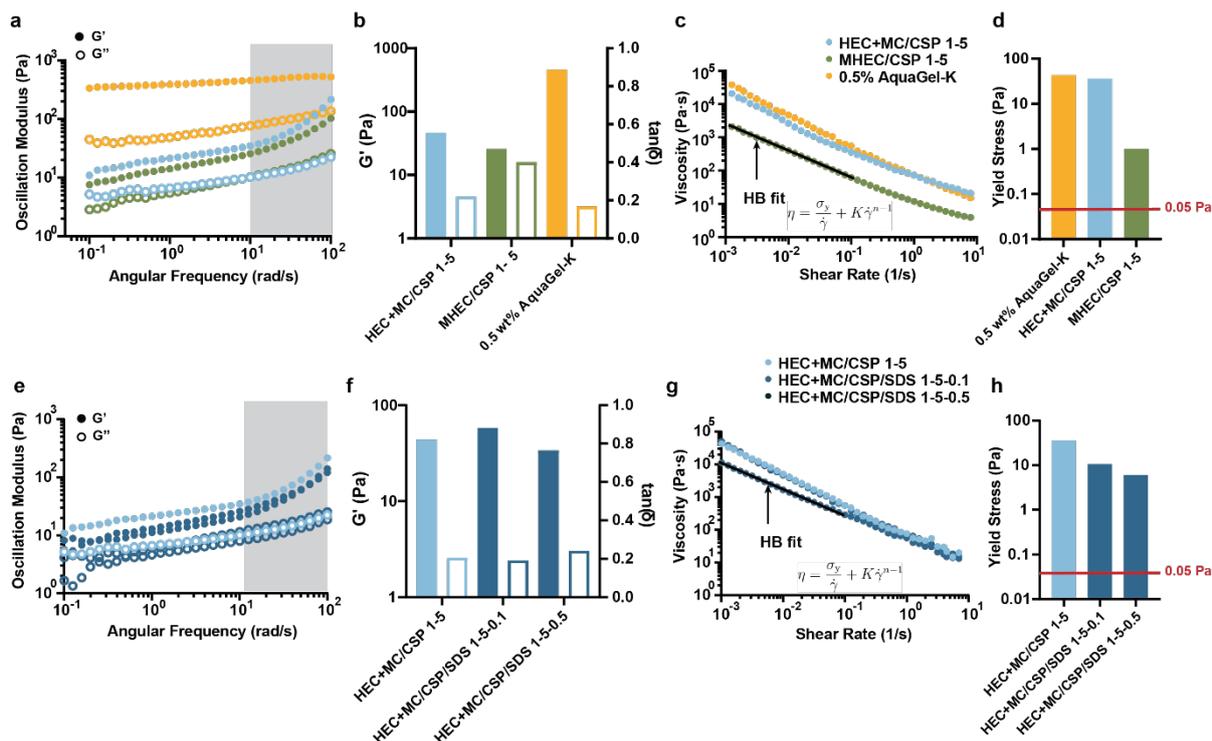

**Figure 2. Tunable shear rheological properties of two distinct non-covalently crosslinked hydrogels. a.** Storage ($G'$) and loss ($G''$) moduli, **b.** $G'$ (left axis) and tan($\delta$) at 0.1 rad/s, **c.** dynamic flow sweep viscosity data, fit to Herschel–Bulkley (HB) model, and **d.** dynamic shear yield stress for two PP hydrogels HEC+MC/CSP 1-5, MHEC/CSP 1-5, and commercial water-enhancing Phos-Chek® AquaGel-K (0.5%) gels. **e.** Storage ($G'$) and loss ($G''$) moduli, **f.** $G'$ (left axis) and tan($\delta$) at 0.1 rad/s, **g.** dynamic flow sweep viscosity data, fit to HB model, and **h.** dynamic shear yield stress for PP hydrogel with three surfactant levels. The grey region in **a** and **e** indicate the region of unreliable data owing to instrument artifacts stemming from rheometer geometry inertial effects.

To confirm the gelation process upon addition of CSPs to the MHEC polymer solutions, we compared the rheology of HEC+MC/CSP 1-5 and MHEC/CSP 1-5 hydrogels with their 1 wt% polymer solutions without CSPs (**Figure S1**). The frequency sweep data test in the top row reveals that the addition of CSP transforms the polymer solution into a gel. The 1 wt% polymer solutions exhibited more fluid-like behavior at low frequencies ($G'' > G'$), moving toward a more solid-like state at higher frequencies ($G' > G''$), whereas the polymer/CSP 1-5 hydrogels consistently showed solid-like behavior with $G' > G''$ across the entire frequency range of 0.1-100 rad/s. The decrease in dynamic moduli observed in some softer samples after 10 rad/s is likely due to instrument artifacts, primarily rheometer geometry inertial effects. The frequency around the modulus crossover serves as an indicator of the viscoelastic network relaxation time ($t_{\text{relax}} \simeq 1/\omega_{\text{cross}}$), while the plateau modulus signifies a pseudo-equilibrium (rubber-like) state within the crosslinked network.[34] These parameters are sensitive to the degree of crosslinking, making frequency sweep tests an efficient method for macroscopic assessment of gelation occurrence. With increasing crosslinking density, the crossover point progressively shifts to lower frequencies until it becomes indiscernible within the tested frequency range.[33] Upon the addition of CSPs to the 1 wt% polymer solutions, the materials undergo a distinct transition to a solid-like hydrogel state, as evidenced by the absence of a crossover point in the 0.1-100 rad/s frequency range we observed, thus indicating the development of long network relaxation times. This observation underscores the material transformation into a PP hydrogel following mixing of MHEC polymers and CSPs, since hydrogels have long relaxation times by definition.[34]

In addition to the initial MHEC derivative we evaluated (supplied by Sigma Aldrich), we prepared PP hydrogels with Walocel® MKX 15000 PP20 and Walocel® MKX 70000 PP01, denoted here as $MHEC_A$ and $MHEC_B$, to further probe the impact of MHEC chemical composition (i.e., degree of substitution of the hydroxyethyl and methyl moieties) on PP hydrogel properties. As $MHEC_A$ has a significantly higher molecular weight than $MHEC_B$, the PP hydrogels prepared with $MHEC_A$ exhibit higher stiffness values and lower tan($\delta$) values. Overall MHEC and $MHEC_A$ based PP hydrogels exhibited very similar properties, so we conducted the remainder of our studies with MHEC-based hydrogels (MHEC/CSP 1-5).

While the PP hydrogels themselves have exceptional potential to perform well as water-enhancing gels, we hypothesized that the addition of surfactants such as sodium dodecyl sulfate (SDS) and tween 20 may improve the fire suppressant behaviors of the hydrogels (**Figure 2e, f**; **Figure S2a**). When PP hydrogels were formulated with surfactants, no significant syneresis was observed with additives at concentrations between 0.1 to 0.5 wt%. Yet, using SAOS frequency sweeps we observed that HEC+MC/CSP 1-5 ($G'$ = 43.9 Pa, tan($\delta$) = 0.207) was stiffer and more elastic-like than the hydrogels formulated with SDS, including HEC+MC/CSP/SDS 1-5-0.1 ($G'$ = 58.2 Pa, tan($\delta$) =0.193) and HEC+MC/CSP/SDS 1-5-0.5 ($G'$ = 33.8 Pa, tan($\delta$) =0.242). These data indicated that addition of surfactant above 0.1 wt% softens the gels (lower values of $G'$) and makes them more dynamic (**Figure 2e, f**).

Beyond the frequency-dependent viscoelastic properties and flow behaviors, the dynamic yield stress serves as crucial property for materials of this type.[20] The dynamic yield stress is an estimate of the threshold stress required to maintain the flow of the material once already in a fluid-like or "yielded" state, providing an indication of the stress needed to keep the hydrogel "fluidized" and thus "sprayable".[35, 36] This property is obtained by fitting the downward flow sweep data (decreasing steady shear rate from 100 to 0.01 s$^{-1}$ to the Herschel-Bulkley (HB) model (**Table S1**).[37] On the other hand, the static yield stress is obtained as the stress at the crossover between the first harmonic storage ($G_1'$) and loss ($G_1''$) moduli during a large amplitude oscillatory shear (LAOS) strain amplitude sweep test, and quantifies the stress required to initiate material flow.[35, 36] The static yield stress is also referred to as the *absolute* yield stress, signifying that materials consistently exhibit fluid-like behavior above this stress threshold and solid-like behavior below it, as observed in LAOS experiments (**Figure S1**). A higher static yield stress indicates a stronger adherence to elevated and vertical fuel surfaces and a greater reluctance to initiate flow. Moreover, the magnitude of strain amplitude at the modulus crossover, often called the yield strain, is an indicator of gradual or abrupt yielding of the material. If the yield strain is small, the material exhibits brittleness. Conversely, if the yield strain is substantial, the material shows stretchiness and a transition to nonlinear flow at larger deformations, suggesting a more gradual yielding. To define the sprayable region for practical applications, one should ensure that the oscillation strain falls within a reasonable range for a spray nozzle. For practical purposes, it is essential to maintain a reasonably high static yield stress while also keeping the dynamic yield stress within a reasonable range suitable for a common spray nozzle design owing to pumping pressure constraints. A comparison of the static and dynamic yield stresses for each formulation is presented in **Figure S3**. The gelation behavior contributes both to a higher viscosity over the shear rate curve and resulting in a higher yield stress for 1-5 PP hydrogel compared to their respective cellulosic polymer solutions (**Figure S1**). Furthermore, the increased strain observed at the crossover point indicates that the hydrogels become "stretchier" upon gelation (**Figure S1**). Careful control and design of these properties in a hydrogel material is essential in obtaining optimal spray, adherence, and coating performance.

## 2.3. Mitigation of Aging Phenomenon of Water-enhancing Gels

The benefits of replacing HEC+MC polymers with MHEC in the PP hydrogel formulation is that the MHEC/CSP system is less susceptible to irreversible chemical aging effects, during which the hydrogels stiffen over time. Initially, the dynamic PP hydrogel forms when HEC and MC polymer chains are adsorbed onto the CSPs, which function as transient noncovalent crosslink junctions between polymer chains. Over time, a fraction of the dynamic, non-covalently bonded crosslinks is converted into covalently bonded anchors, making these junctions irreversible chemical bonds (**Figure S4**). The hydrogel becomes stiffer with age and shows higher storage and loss moduli as confirmed by rheology experiments (**Figure S5**). The stiffening phenomenon is due to the strengthening of existing non-covalent binding sites and preserving the overall density of crosslinks, as opposed to increasing crosslink density by forming new junctions. The presence of the hydroxyl (–OH) groups on the adsorbed cellulose chains and the silanol (Si–O–H) groups on the surface of CSPs result in a condensation reaction leading to the formation of siloxane (Si–O–Si) bonds, supported by stress-relaxation rheology and FTIR-ATR characterization across an aging span of 15 days (**Figure S6** and **S12**).[26] Both HEC and MC have an abundance of free –OH groups on the sugar rings along the polymer backbone that induce the noncovalent-to-covalent transition in a significant fraction of the PP crosslinks. We hypothesized that using cellulose variants with a smaller fraction of free –OH groups could prevent these aging effects. MHEC was selected based on its backbone that contains fewer free –OH groups available for reaction, since the sugar moieties are modified with both hydroxyethyl and methyl groups. Moreover, the requirement for only two components in MHEC-based hydrogels (MHEC and CSP) rather than three components (HEC, MC, and CSPs) simplifies manufacturing. Furthermore, MC is more expensive and more susceptible to contamination than MHEC and HEC derivatives, so removing it from the formulation can also reduce costs of these materials. The common applications of Walocel® MHEC include spray plaster and cement-based tile adhesives, and compatible with all conventional mineral and organic binders, thus not compromising with the versatility and compatibility achieved with the HEC+MC polymers.

## 2.5. Burn Tests and Aerogel Formation upon Heat Activation

The fire retarding effect of water-enhancing gels upon heat activation was evaluated by the duration of direct exposure to a flame from a Benzomatic MAP Gas Hand Torch. MAP-Pro® gas is a stabilized mixture of propylene and propane that can burn at up to 2054 °C at 1 atm.[38] In these studies, the different hydrogel formulations discussed earlier were applied to a plywood plate in a uniform layer (~1/4 inches thick) and subjected to direct flame exposure until the wood substrate began to char (**Figure 3a**). The time-to-char was used as a metric of the fire-retardant performance. The formulations most effective at preventing charring were HEC+MC/CSP 1-5 and HEC+MC/CSP/SDS 1-5-0.1, protecting the substrate for more than 7 min of continuous flame exposure. Similarly, MHEC/CSP 1-5 PP formulations protected the substrates for over 5 min. Almost immediately upon direct flame impingement, the PP hydrogels form a solid opaque crust that scatters the heat from the flame. Over time with continued flame exposure, the PP materials form a foam, desiccate, and form a silica aerogel *in situ* that increasingly insulates the coated substrates from the flame. In contrast, the AquaGel-K gels protected the substrates from charring for less than 90 seconds, when the water encapsulated in these gels locally desiccate, demonstrating how these traditional materials only function while wet. These observations indicate that PP hydrogel formulations are significantly more effective as water-enhancing gels than the leading commercial product (n = 3-6 for all tests; **Figure 3c**). Videos of the flame exposure on MHEC/CSP 1-5 and AquaGel-K treated substrates are reported in the supplementary information (**Video S1** and **S2**).

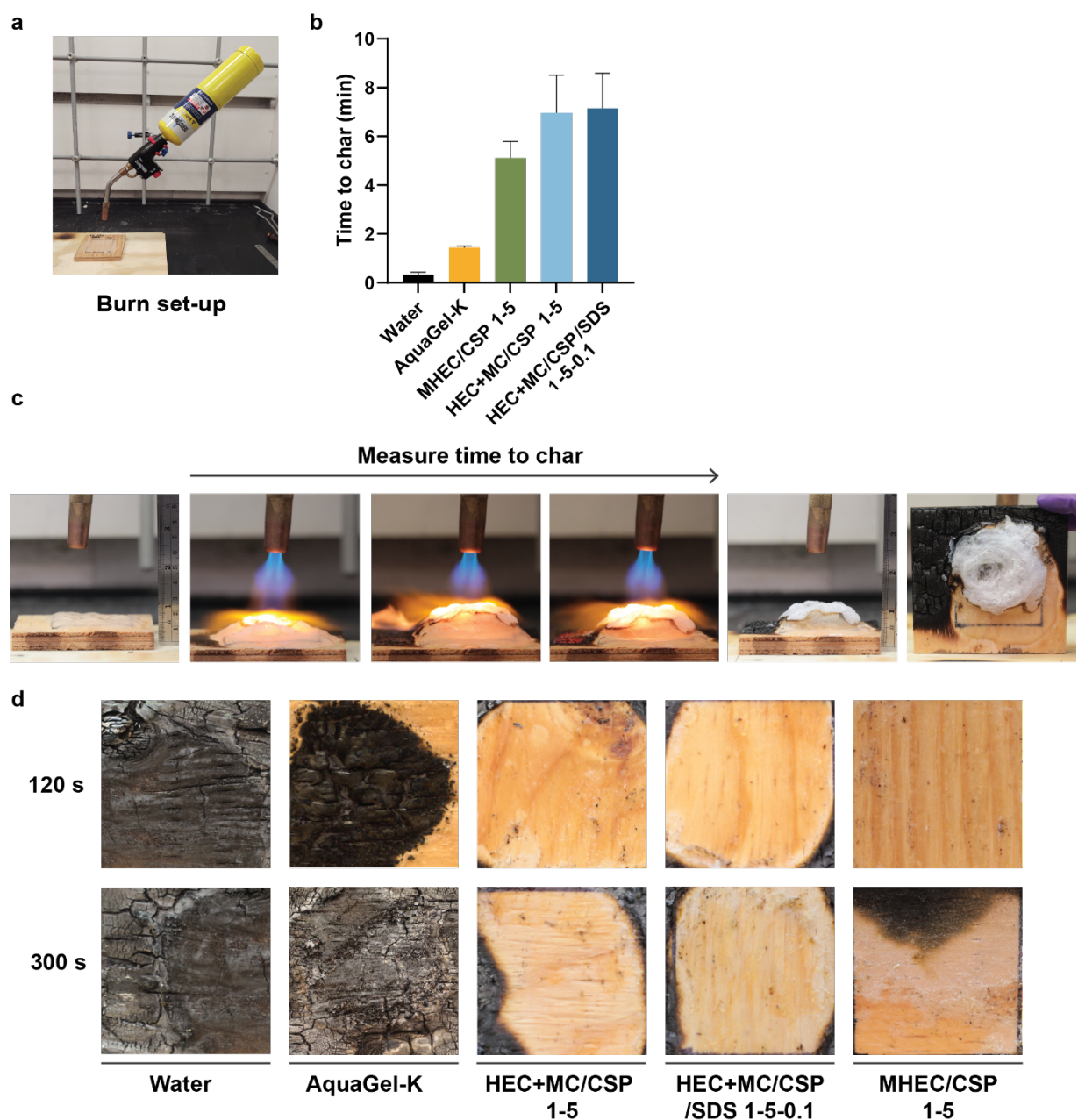

**Figure 3**. **PP hydrogel rheological properties impact burn performance. a.** An image of the burn set-up. **b.** Time taken to char the wood underneath the surface treatment. **c**. Images of the burn process for a MHEC/CSP 1-5 hydrogel. **d.** Images of burnt wood under treatments of water, AquaGel-K, HEC+MC/CSP 1-5, HEC+MC/CSP/SDS 1-5-0.1, and MHEC/CSP 1-5 at burn times of 120 s and 300 s.

Aerogels, which are typically mesoporous with high specific surface area (500 ~ 1000 m$^2$/g) and low density (0.001 ~ 0.200 g/cm$^3$), are classified as super-insulators due to their ultralow thermal conductivity (as low as 12 mW/m/K).[39, 40] To evaluate and quantify the foaming and silica aerogel formation process for the PP hydrogels under heat activation, we defined a "foaming index" based on the film thickness pre- and post-burn. The foaming film thickness was measured as the average distance between the top surface of the aerogel and the surface of the wood substrate (**Figure 4a, S7a**). The foaming index was defined as the ratio of the final aerogel film thickness to the initial hydrogel film thickness and compared across the four major formulations (**Figure 4b, S7b**). We see that both HEC+MC/CSP 1-5 and MHEC/CSP 1-5 materials, and their derivative formulations, are excellent at protecting the substrate, as exhibited by the extensive time-to-char (**Figure S8**). When comparing the time-to-char to the foaming index, a general trend is observed whereby higher foaming indexes result in better substrate protection. Interestingly, we had hypothesized that inclusion of surfactants would increasing foaming, yet the HEC+MC/CSP 1-5 formulation exhibits the highest foaming index of all the formulations evaluated. This observation may be explained by the fact that increasing surfactant loading enlarges the pores of the resulting aerogel foams, potentially leading to the formation of wider gaps between stacked aerogel layers, which was confirmed by scanning electron microscopy (SEM) imaging of the foams (**Figure 4c**). Microscopically, the silica particles were not affected by addition of surfactant, yet the overall aerogel morphology was altered. There were no observable morphological differences between aerogel films formed with PP hydrogels comprising MHEC and HEC+MC polymers. Moreover, PP hydrogel formulations comprising the two Walocel® MHEC derivatives ($MHEC_A$ and $MHEC_B$) exhibited comparable aerogel and particle morphologies as the MHEC/CSP 1-5 formulation, as confirmed by SEM micrographs (**Figure S9**).

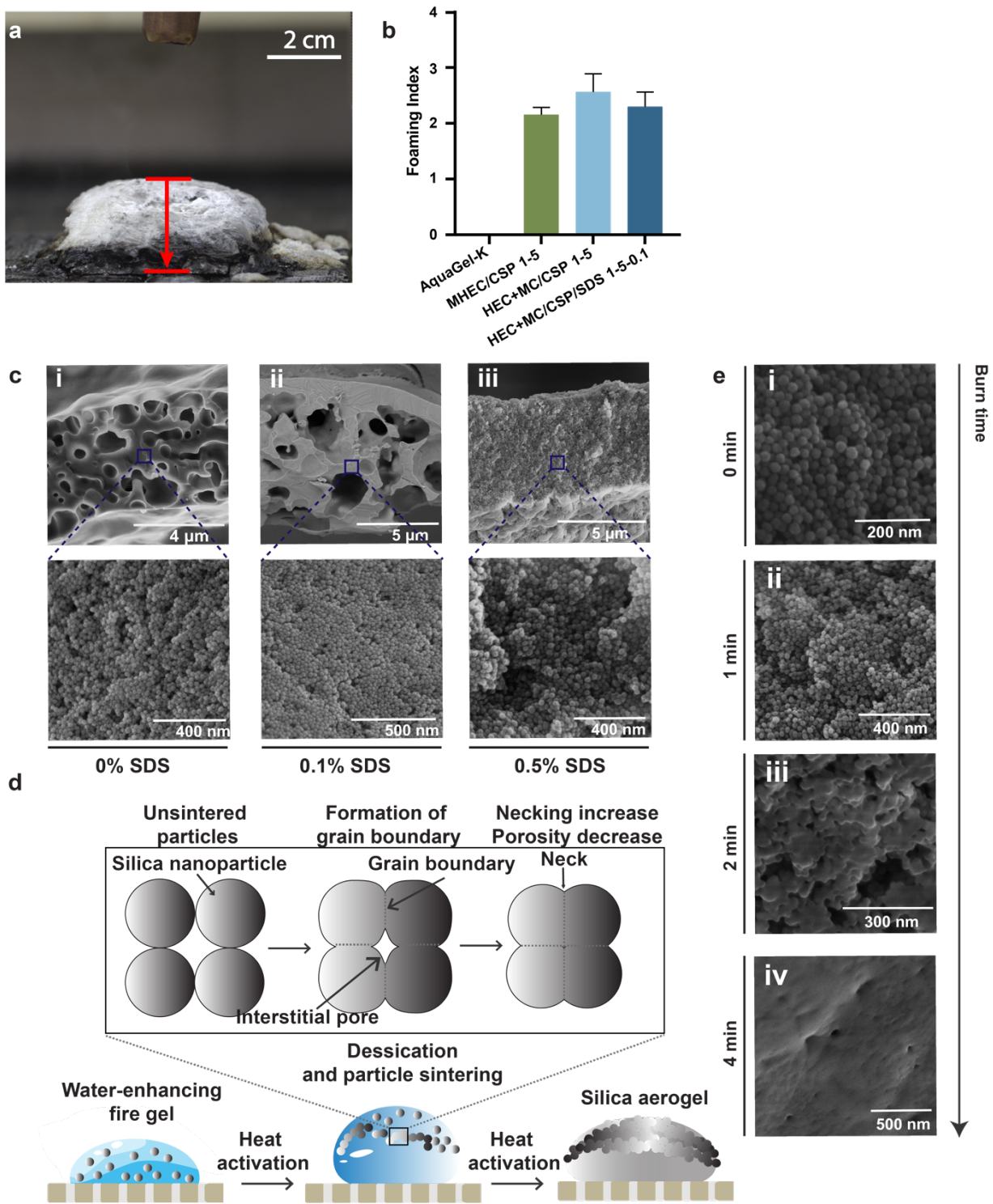

**Figure 4. Silica particle densification aids PP hydrogel fire retardancy. a.** Post-burn final film thickness was measured as the vertical distance between the top of the wood substrate and the top of the aerogel canopy. **b.** Foaming index (the ratio between the aerogel film after burn and the hydrogel film thickness before burn) for HEC+MC/CSP 1-5, HEC+MC/CSP/SDS 1-5-0.1, MHEC/CSP 1-5, and AquaGel-K. **c.** SEM characterization of the cross-section view (top row) and higher magnification view of CSP (bottom row) for HEC+MC/CSP 1-5, HEC+MC/CSP/SDS 1-5-0.1, and HEC+MC/CSP/SDS 1-5-0.5. **d.** Schematics for particle sintering under heat and transition from hydrogel to aerogel during heat activation. **e.** SEM of CSP for film samples subjected to different burn times.

The mechanism of protection of the wood substrate for AquaGel-K was soley based on retention of water within the hydrogels and all the solid content of the AquaGel-K materials was consumed during the burn. In contrast, all PP hydrogel materials formed a robust silica aerogel foam during the burn whereby all the cellulose content was burned away, as observed by FT-IR analysis of the aerogels (**Figure S11** and **S12**). Moreover, while at early timepoints of flame exposure (less than 2 min) the silica particles maintain their morphology, at longer timepoints (greater than 2 min) the silica particles begin to sinter, leaving a solid silica foam (**Figure 4d** and **4e**). This morphology provides excellent insulation, scattering heat from the flame and protecting the wood substrates from charring for several more minutes of continuous flame exposure. These materials thus provide a robust mechanism for thermal insulation and char protection compared to commercial water-enhancing hydrogel materials.

## 3. Conclusion

Considering the escalating wildfire crisis, we urgently need innovative solutions for wildfire prevention and protection. Water-enhancing gels are promising tools for fire suppression due to their remarkable wetting and favorable viscoelastic properties, which play a pivotal role in reducing the risk of ignition and enhancing surface adherence. Unfortunately, a significant drawback lies in the fact that traditional water-enhancing hydrogels lose their effectiveness when subjected to the high heat and strong winds that are commonplace during wildfires, thereby limiting their practical utility dramatically. To address this critical limitation, our research has introduced a novel heat-activatable hydrogel design leveraging dynamic, multivalent polymer-particle interactions between cellulosic biopolymers and colloidal silica particles that can be fabricated in a facile and scalable manner. When exposed to heat, these hydrogels undergo a transformation as the gel desiccates from applied heat, converting *in situ* into a thermally insulative aerogel. This innovative feature magnifies the protective capabilities of these materials as coatings on substrates, ensuring continued efficacy against ignition even when desiccated. The use of cellulose derivatives, based on the most abundant biopolymer on earth, ensures cost-effectiveness and environmental safety of these materials. The mechanism of strong physical crosslinking between these cellulosic biopolymers and colloidal silica particles offers a wide range of tunability in mechanical properties while enabling exceptional adherence to substrates and superior fire retardancy than traditional water-enhancing gels. This versatility positions these materials as extraordinarily effective water-enhancing gels. Through tuning of the polymer-particle interactions, we could overcome challenges with irreversible aging observed for some existing hydrogel systems, enabling formulation of materials with enhanced properties for practical usage. These innovative, bioinspired, and environmentally benign materials, have the potential to safeguard lives, property, and the environment from extreme wildfire.

## 4. Experimental Section
### 4.1 Materials
Hydroxyethyl methyl cellulose products Walocel® MKX 15000 PP20 and Walocel® MKX 70000 PP01 were provided by Dow® Chemical Company. All other materials were purchased from Sigma Aldrich and used as received.

### 4.2. Methods
*PP Hydrogel Preparation*

Hydrogel formulations were prepared by following published protocol for making HEC+MC/CSP viscoelastic fluids reported in our previous study in a ratio of 0.8/0.2/5 wt% in water.[21] Polymer solutions were first prepared by dissolving hydroxyethyl cellulose (HEC, $M_v$ ≈ 1,300 kDa, Sigma Aldrich) and methylcellulose (MC, $M_v$ ≈ 90 kDa, Sigma Aldrich) in water at a concentration of 30 mg/mL. The polymer solution was stirred vigorously overnight at room temperature for full dissolution. CSP used in the study is LUDOX® TM-50 (Sigma Aldrich), a dispersion of amorphous anionic silica with a particle size of 22 nm in diameter produced by polarizing silica nuclei from silicate solutions under alkaline conditions. CSP is often used as a frictionizing agent in paper and textile coatings. To achieve uniform dispersion in gel formulation, CSP was diluted from 50 wt% to 15 wt% (pH = 9) with water and slowly added dropwise into the HEC+MC polymer solution while stirring with a spatula at a 1:5 ratio by weight. Extra water was added to reach the final concentration of 1-5 gel. For the formulations with surfactant, sodium dodecyl sulfate (SDS) or Tween 20 were added to the polymer solution before adding CSP.

*Rheological characterization:*

The rheological properties of the hydrogels were measured using a 20 mm serrated parallel plate on a stress-controlled TA Instruments Discovery HR-2 rheometer at a gap of 750 $\mu$m. The 1 wt% mixture of polymer solution in the SI was tested using a 60 mm cone and plate geometry on the same rheometer. All tests were carried out at 25 °C. Frequency sweep tests were performed by applying an oscillatory strain signal input with frequencies of 0.1 to 100 rad/s at 1% strain under controlled normal force condition with 10 points per decade. Flow sweeps were performed at shear rates of $10^1$ to $10^{-6}$ 1/s with steady-state detection. Large Amplitude Oscillatory Shear (LAOS) tests were performed at a frequency of 1 rad/s over strain amplitudes of 0.1 to 10,000% with 4 points per decade. Stress relaxation tests for HEC+MC/CSP and MHEC/CSP were performed on a separated motor-transducer rheometer (ARES-G2, TA Instruments) at Stanford Soft and Hybrid Materials Facility (**Figure S5**). The Herschel-Bulkley (HB) model (defined as $\sigma(\dot{\gamma}) = \sigma_y + K\dot{\gamma}^n = \eta(\dot{\gamma})\dot{\gamma}$, where $\sigma_y$, $K$, and $n$ are the model parameters) was used to obtain the dynamic yield stress $\sigma_y$ by fitting the flow sweep data in MATLAB. Static yield stress was obtained from the first harmonic storage ($G_1'$) and loss ($G_1''$) moduli crossover from a strain-controlled LAOS test based on linear/cubic spline combination method in TA Instruments TRIOS software.

*Burn Tests:*

Hydrogel (20g) was applied onto a plywood plate in a square area 2.54 cm × 2.54 cm (Lowe's® Unfinished Whitewood Board, cut into squares of 5.08 cm × 5.08 cm). Benzomatic® MAP-Pro® Premium Hand Torch Fuel based on propylene (99.5 – 100 vol%) and propane (0 – 0.5 vol%), which can burn up to 2054 °C (3730 °F),[38] was used for burns. The torch was stabilized ~ 5 cm above the hydrogel surface. Burn time was taken between the time when the hand torch was turned on and the time when the wood substrate started to visually char. The distance between the top of the aerogel foam and the surface of the wood plate was taken as foam thickness. The aerogel foam was collected post burn to conduct electron microscopy analysis for imaging the foam structure and measuring the film thickness.

*Scanning Electron Microscopy:*

The morphology and thickness of the burnt aerogel films were characterized by scanning electron microscopy (SEM) using the FEI Magellan 400 XHR Scanning Electron Microscope at Stanford Nano Shared Facility (SNSF). The samples were grounded to a 90-degree aluminum pin stub using double-sided conductive copper tape. 2 ~ 10 nm thick layer of Au:Pd (60:40) was sputtered on the samples to minimize charging prior to imaging. The imaging analysis was performed at a beam voltage of 5.00 kV, and high vacuum in immersion mode.

**Data Availability Statement**

The data that support the findings of this study are available on request from the corresponding author.

**Supporting Information**

Supporting Information is available from the Wiley Online Library or from the author.

**Conflicts of Interest Statement**

E.A.A. is an inventor on a patent that describes the technology reported in this manuscript. All other authors declare no conflicts of interest.

**Acknowledgements and Author Contributions**

C.D., A.I.D., and E.A.A. conceived of the idea and experimental planning. C.D. and A.I.D. performed experiments. Other authors aided with experiments. We appreciate the Gordon & Betty Moore Foundation for their financial support of this work as part of our efforts to develop wildland fire retardants to improve the execution of prescribed burns and enhance forest management. A.I.D. was supported by a Schmidt Science Fellows Award. Part of this work was performed at the Stanford Nano Shared Facilities (SNSF), supported by the National Science Foundation under award ECCS-2026822. Part of the rheology data was collected at the Stanford Soft and Hybrid Materials Facility (SMF) (supported by the NSF grant National Science Foundation grant ECCS1542152).

Received: ((will be filled in by the editorial staff))
Revised: ((will be filled in by the editorial staff))
Published online: ((will be filled in by the editorial staff))

Supplementary Information

# Water-enhancing gels exhibiting heat-activated formation of silica aerogels for protection of critical infrastructure during catastrophic wildfire

*Changxin Dong, Andrea I. d'Aquino, Samya Sen, Ian A. Hall, Anthony C. Yu, Jesse D. Acosta, and Eric A. Appel\**

# Table of Contents

**Supplementary Table**

**Table S1.** List of MHEC polymers used in this study, their suppliers, and properties.

| Name | Product | Supplier | Properties | Applications |
|---|---|---|---|---|
| MHEC | Methyl 2-hydroxyethyl cellulose ($C_{34}H_{66}O_{24}$) | Sigma-Aldrich | 858.9 g/mol, Viscosity 15000 - 20500 cps, c = 2%, Water, 20°C | Water retention aid, thickening agent, protective colloid, suspending agent, binder and stabilizer. |
| $MHEC_A$ | WALOCEL™ MKX 15000 PP 25 | The Dow Chemical Company | Viscosity 15000 cps, c = 2%, Water, 20°C | Cement spray plaster, EIFS |
| $MHEC_B$ | WALOCEL™ MKX 70000 PP01 | The Dow Chemical Company | Viscosity 9900 cps, 1%; aqueous solution | Cement-based tile adhesives |

**Supplementary Figures**

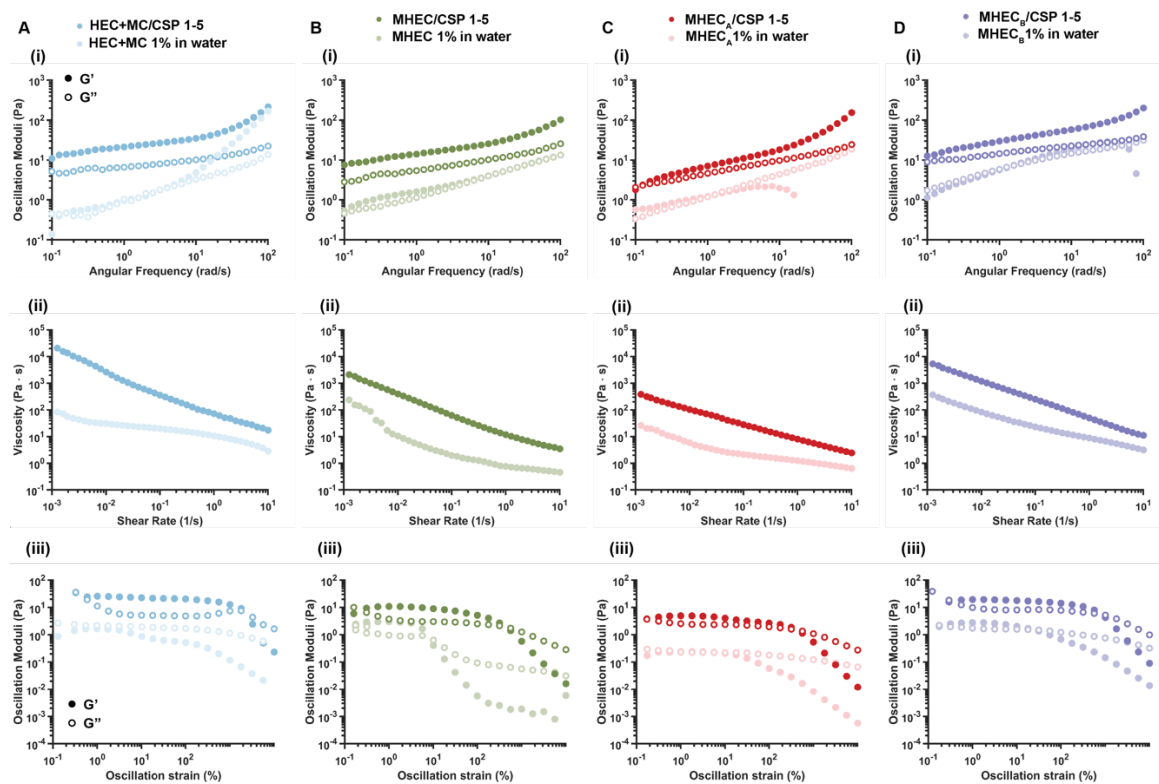

**Figure S1. Rheology of polymer-particle hydrogel formulations.** Experiments include: (i) strain-dependent oscillatory shear measurements ($\omega$ = 10 rad/s, 25 °C), (ii) frequency-dependent oscillatory shear measurements ($\varepsilon$ = 1%, 25 °C), and (iii) steady shear measurements (25 °C). **a.** HEC+MC/CSP 1-5 (darker blue) and 1 wt% HEC+MC in water (lighter blue). **b.** MHEC/CSP 1-5 (darker green) and 1 wt% MHEC in water (lighter green). **c.** MHEC$_A$/CSP 1-5 (darker red) and 1 wt% MHEC$_A$ in water (lighter red). d. MHEC$_B$/CSP 1-5 (darker purple) and 1 wt% MHEC$_B$ in water (lighter purple). Grey boxes over frequency-dependent oscillatory shear measurements indicate where inertial effects make data unreliable.

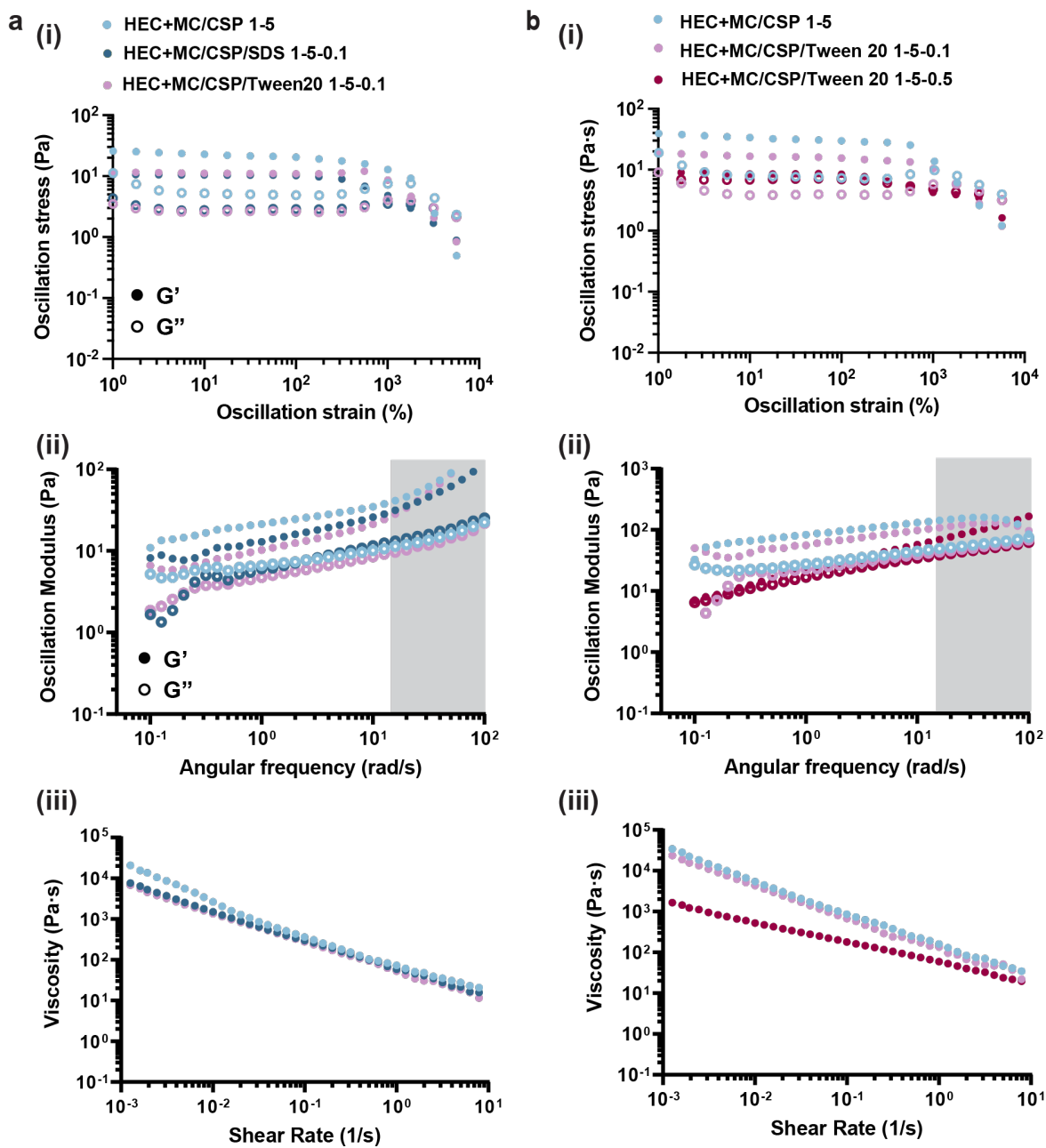

**Figure S2. Rheology for HEC+MC/CSP 1-5 hydrogel with anionic (SDS) and non-ionic (Tween20) surfactant additives.** Experiments include: (i) strain-dependent oscillatory shear measurements ($\omega$ = 10 rad/s, 25 °C), (ii) frequency-dependent oscillatory shear measurements ($\varepsilon$ = 1%, 25 °C), and (iii) steady shear measurements (25 °C). **a.** HEC+MC/CSP 1-5 (light blue), HEC+MC/CSP/SDS 1-5-0.1 (dark blue), and HEC+MC/CSP/Tween20 1-5-0.1 (purple). **b.** HEC+MC/CSP 1-5 (light blue), HEC+MC/CSP/Tween20 1-5-0.1 (purple), HEC+MC/CSP/Tween20 1-5-0.5 (darker purple).

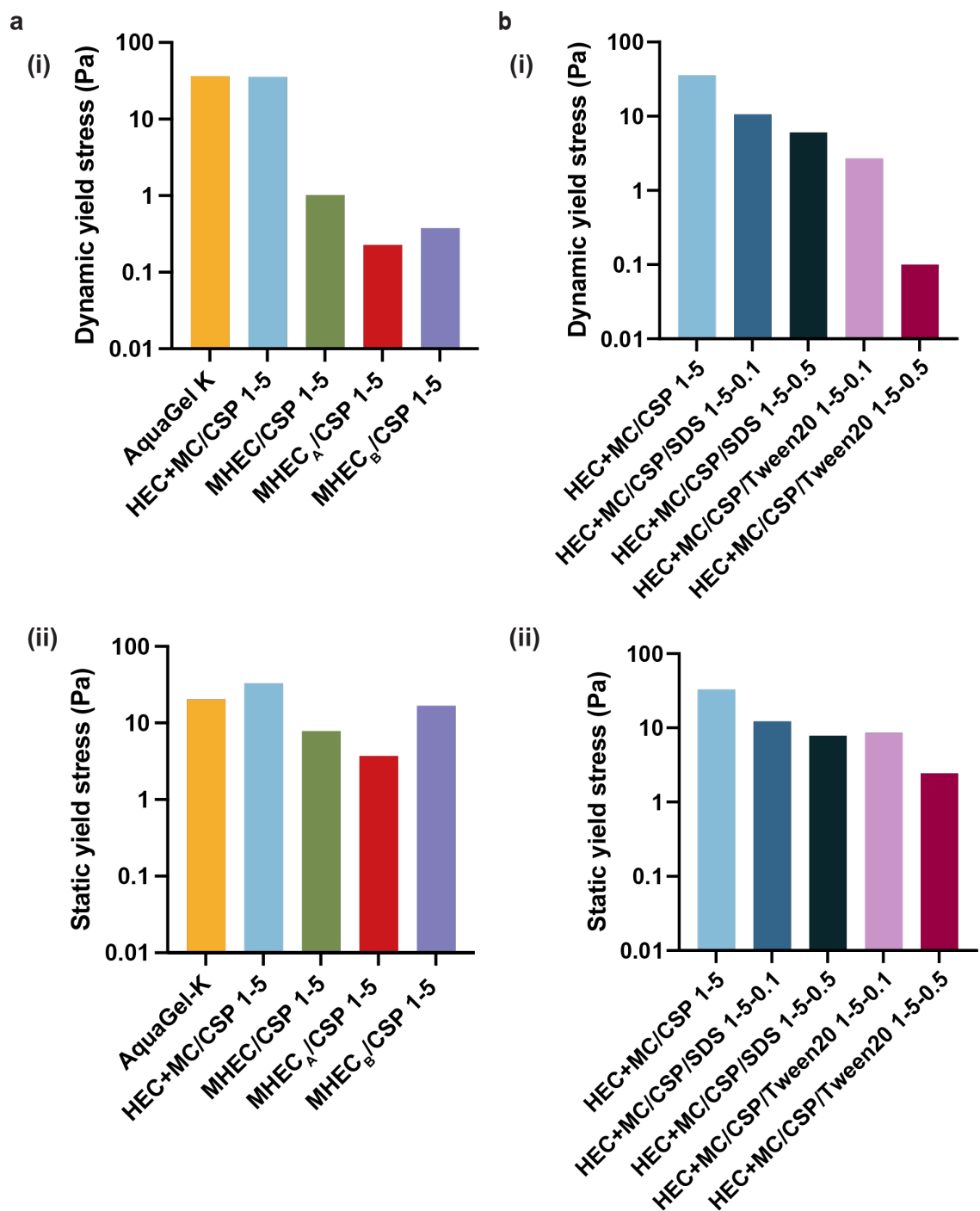

**Figure S3. Yield stress behaviors.** Dynamic yield stress (i) values for various formulations obtained by fitting the HB model to shear rheological data, and static yield stress (ii) values taken at the first crossover on a LAOS test. **a.** AquaGel-K, MHEC/CSP 1-5, HEC+MC/CSP 1-5, MHEC$_A$/CSP 1-5, and MHEC$_B$/CSP 1-5. **b.** HEC+MC/CSP 1-5, HEC+MC/CSP/SDS 1-5-0.1, HEC+MC/CSP/SDS 1-5-0.5, HEC+MC/CSP/Tween20 1-5-0.1, and HEC+MC/CSP/Tween20 1-5-0.5.

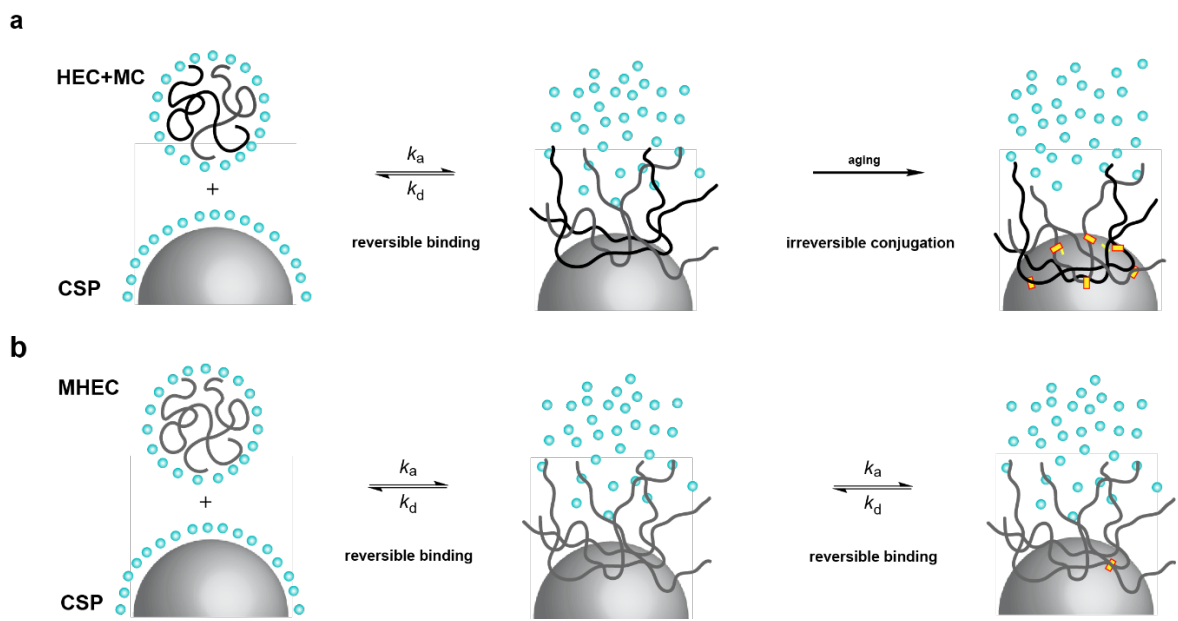

**Figure S4. Schematic of the aging process in polymer-particle hydrogels.** During aging for polymer-particle hydrogels, the interactions transition from reversible binding to irreversible conjugation (indicated by the yellow "anchor" points). Polymers are first absorbed onto the CSP by dynamic, non-covalent interactions to form reversible binding. Over time, MHEC/CSP forms fewer irreversible conjugations compared to HEC+MC/CSP, resulting in more subtle aging behavior. **a.** In HEC+MC/CSP gels, this covalent conjugation process is extensive. **b.** In contrast, in the MHEC/CSP hydrogel system, there are fewer available sites to form the irreversible conjugation.

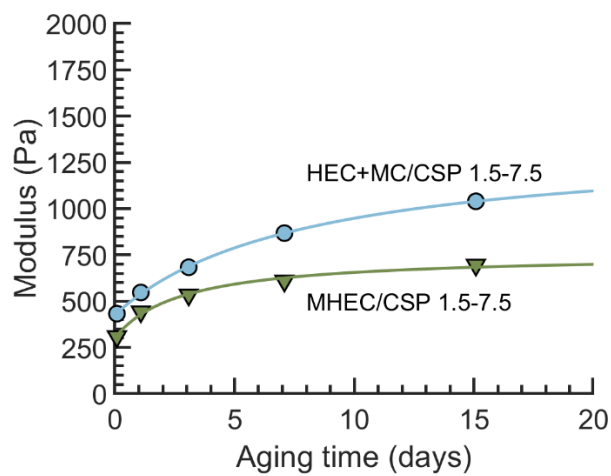

**Figure S5. Mechanical characterization of aging on polymer-particle hydrogels.** The aging effect was observed over the course of 20 days by characterizing the modulus values for HEC+MC/CSP 1.5-7.5 (top) and MHEC/CSP 1.5-7.5 (bottom). The fit lines are based on an model for the second order kinetic reaction underpinning the dynamic-to-covalent transition that was described in a previous study.[1]

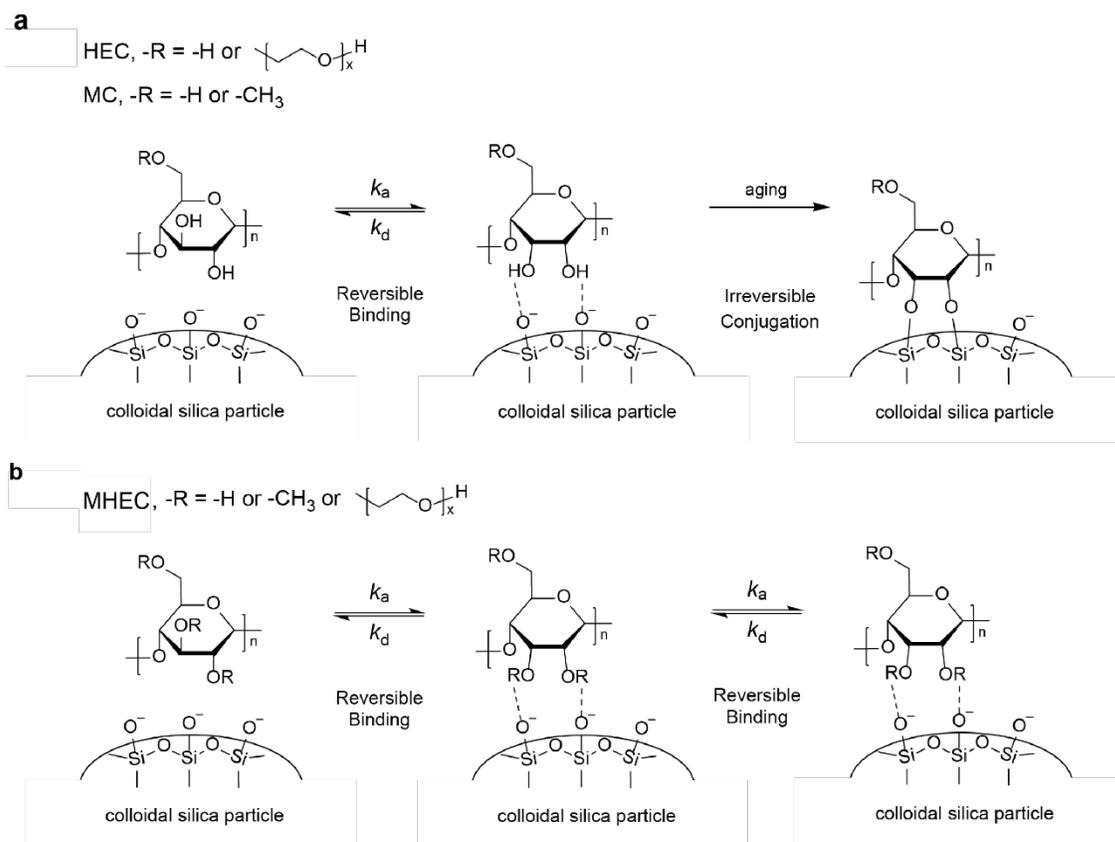

**Figure S6. Surface chemistry for hydrogel aging.** Interfacial interactions between a. HEC+MC chains and the surface of CSP involve an initial reversible interaction of polymer chains with the particle surface, driven in part by hydrogen bonding effects. Over time, the alcohol groups between the HEC+MC chains and CSP undergo irreversible conversion to form covalent and polar silicon-oxygen bonds during a condensation reaction. b. MHEC chains absorb onto the surface of CSP to form reversible interactions. Due to fewer alcohol groups on MHEC compared to HEC and MC, they are less susceptible to forming irreversible junctions. Over time, most bonds remain reversible in the MHEC/CSP hydrogel. $k_a$ is defined as the rate of cross-link association and $k_d$ is defined as the rate of cross-link dissociation.

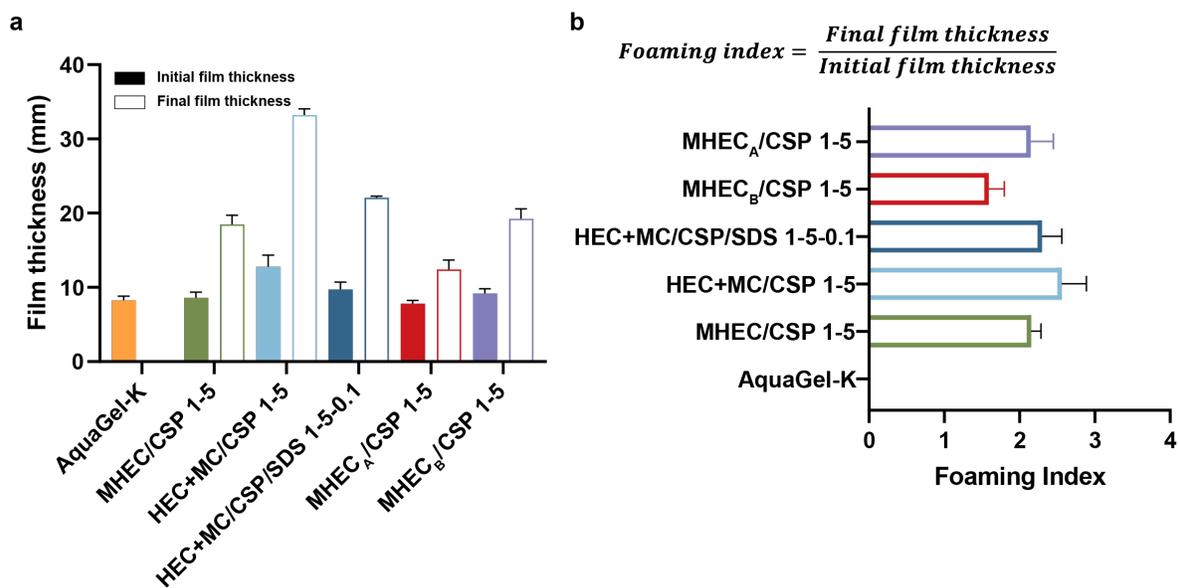

**Figure S7. Quantification of film foaming. a.** The initial and final film thickness for AquaGel-K, MHEC/CSP 1-5, HEC+MC/CSP 1-5, HEC+MC/CSP/SDS 1-5-0.1, $MHEC_A$/CSP 1-5, and $MHEC_B$/CSP 1-5. **b.** The film indices for water, AquaGel-K, MHEC/CSP 1-5, HEC+MC/CSP 1-5, HEC+MC/CSP/SDS 1-5-0.1, $MHEC_A$/CSP 1-5, and $MHEC_B$/CSP 1-5 (bottom to top). Film index is defined as the ratio between the final aerogel film thickness and initial hydrogel film thickness.

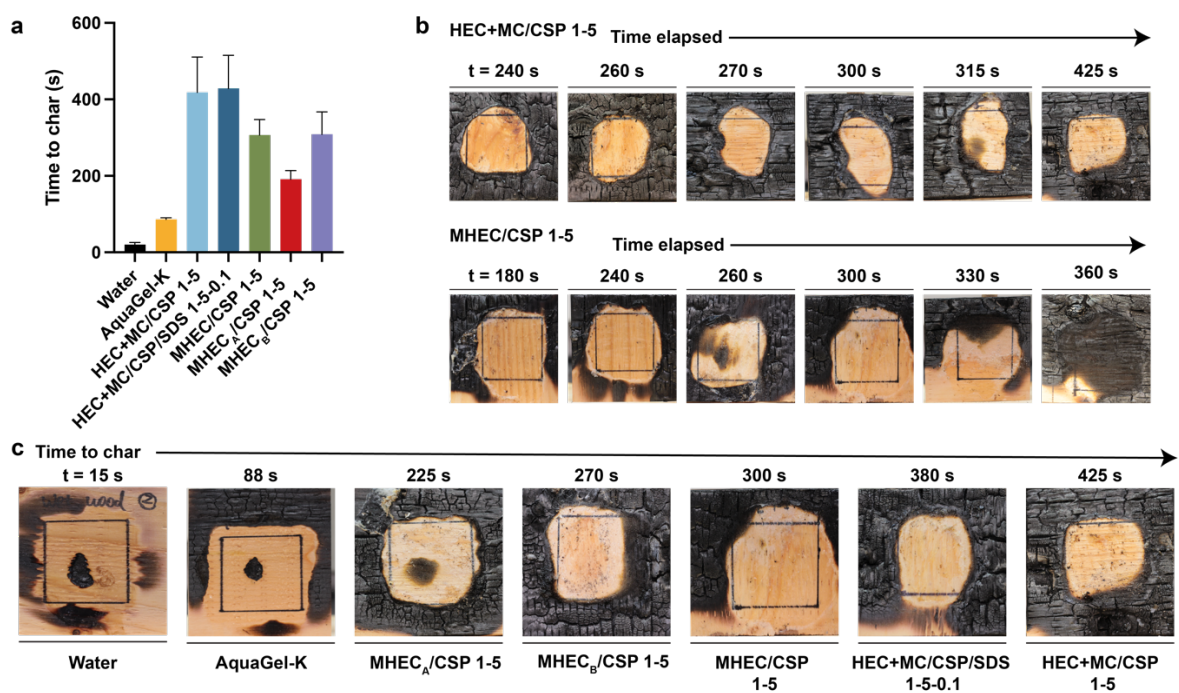

**Figure S8. Protection of substrates from impinging flame. a.** Effectiveness of surface protection was evaluated by the time taken to char the substrate underneath the treatments of AquaGel-K, MHEC/CSP 1-5, HEC+MC/CSP 1-5, HEC+MC/CSP/SDS 1-5-0.1, MHEC$_A$/CSP 1-5, and MHEC$_B$/CSP 1-5 (3≤ n ≤7). **b.** Images of the post-burn wood substrate treated by HEC+MC/CSP 1-5 with burn time = 240 s, 260 s, 270 s, 300 s, 315 s, 425 s; and 180 s, 240 s, 260 s, 300 s, 330 s, and 360 s. **c.** Images of the post-burn wood substrate after cleaning up the surface treatments, ranked by the burn time, in order of the time taken to char (from left to right).

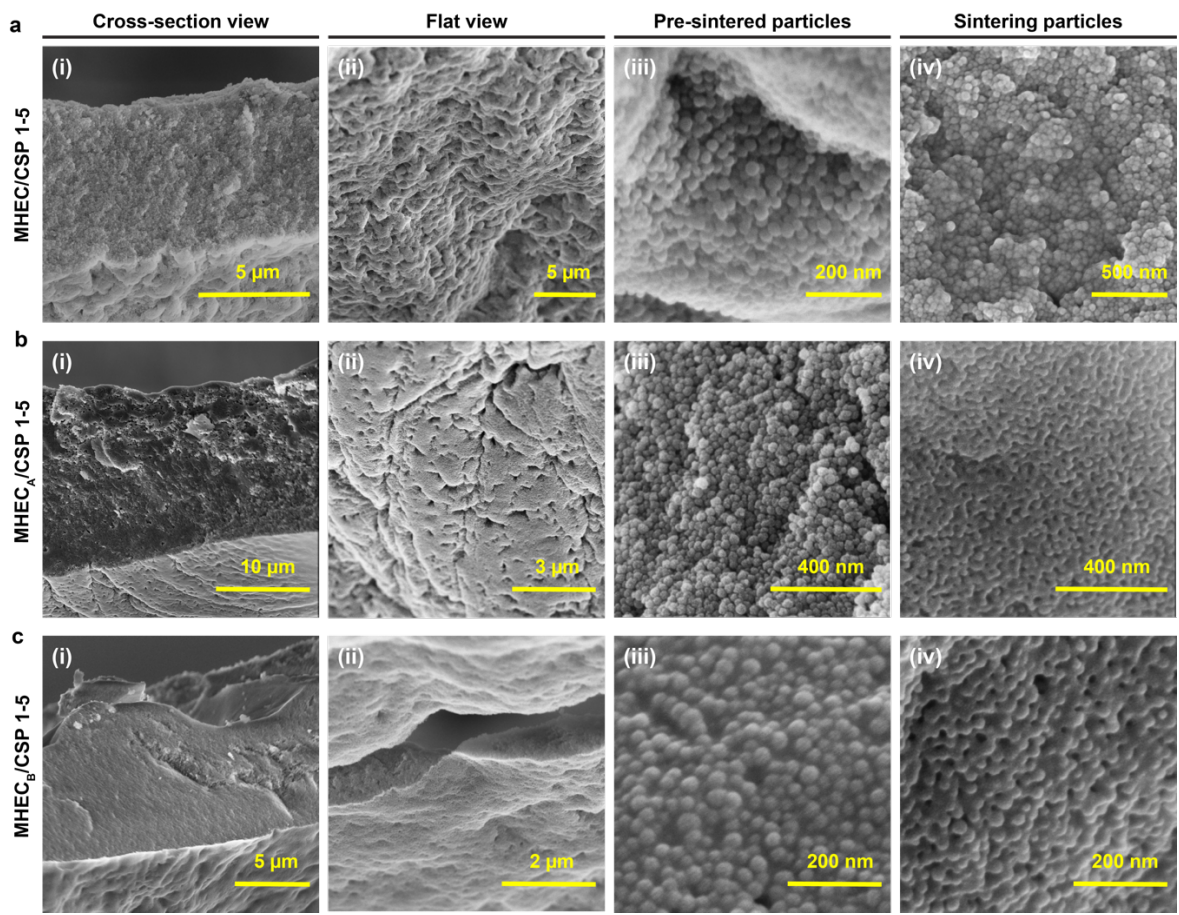

**Figure S9. SEM micrographs for aerogel formed by 3 MHEC hydrogel formulations.** SEM view from (i) cross-section, (ii) flat surface, (iii) pre-sintered silica particles, and (vi) sintering silica particles on the aerogel. Samples include **a.** MHEC/CSP 1-5, **b.** MHEC$_A$/CSP 1-5, and **c.** MHEC$_B$/CSP 1-5.

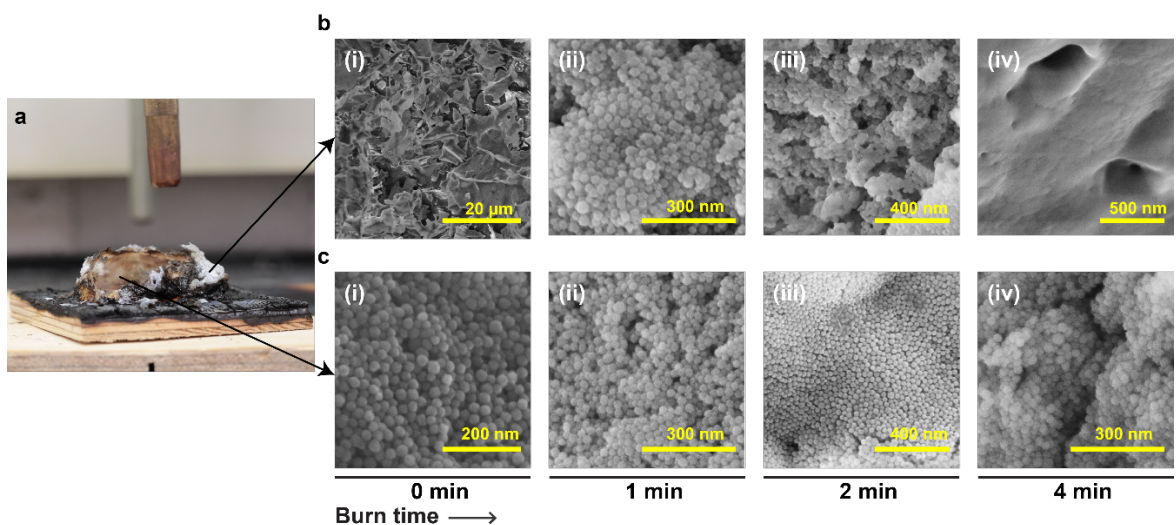

**Figure S10. SEM images for sintering process on the upper and lower layer of the HEC+MC/CSP 1-5 aerogel. a.** Image showing the locations of the aerogel film sampled for SEM. **b.** Sintering process of the aerogel layer closest to the flame at burn time of (i) 0 min, (ii) 1 min, and (iii) 4 min. c. Sintering process of the aerogel layer closest to the wood substrate at burn time of (i) 0 min, (ii) 1 min, and (iii) 4 min.

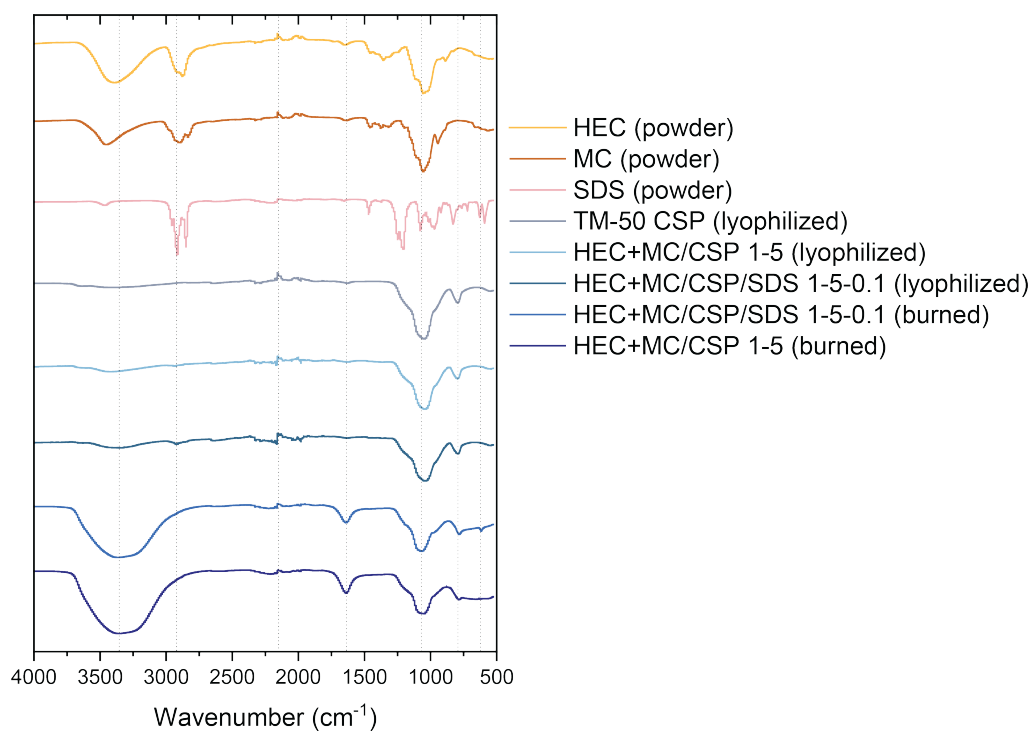

**Figure S11. FT-IR ATR measurements for hydrogel components.** Powdered HEC (yellow), MC (orange), SDS (pink), lyophilized TM-50 silica particles (grey), lyophilized HEC+MC/CSP 1-5 (light blue), lyophilized HEC+MC/CSP/SDS 1-5-0.1 (dark blue), burned HEC+MC/CSP/SDS 1-5-0.1 crushed into powder (sapphire blue), and burnt HEC+MC/CSP 1-5 crushed into powder (purple).

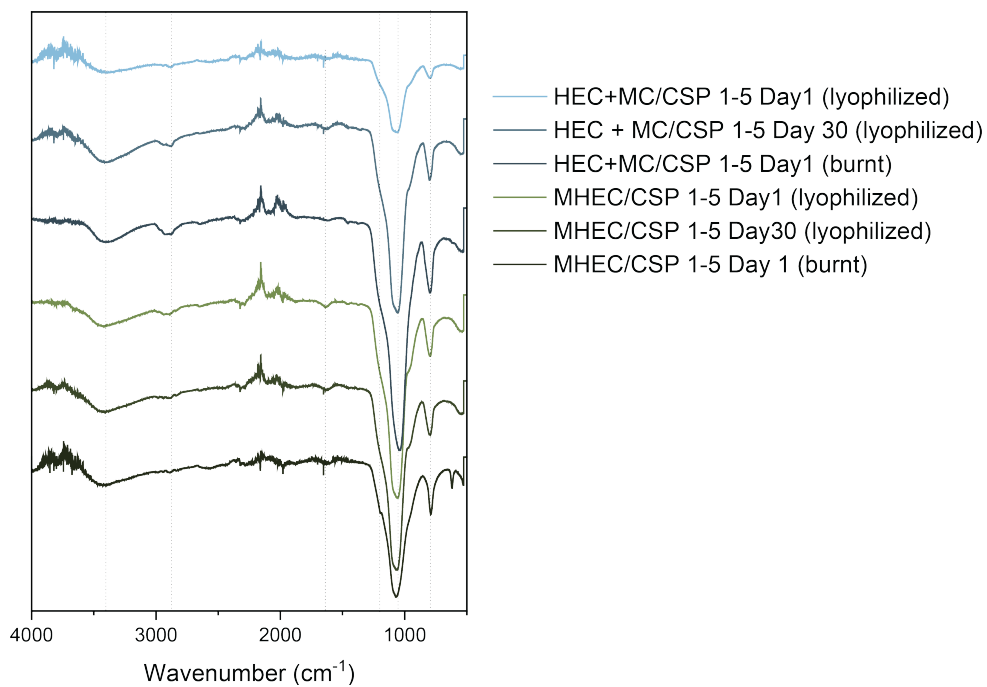

**Figure S12. FT-IR ATR measurements for hydrogels.** Lyophilized HEC+MC/CSP 1-5 Day1, lyophilized HEC+MC/CSP 1-5 Day30, burned HEC+MC/CSP 1-5 crushed into powder, burnt HEC+MC/CSP 1-5 crushed into powder, lyophilized MHEC/CSP 1-5 Day1, lyophilized MHEC/CSP 1-5 Day30, burned MHEC/CSP 1-5 crushed into powder, and burnt MHEC/CSP 1-5 crushed into powder.

**Supplementary Videos**

**Video S1.** Video of MHEC/CSP 1-5 burn played at 3x speed, featuring images of the post-burn silica canopy and the post-burn substrate. See the video at:
https://www.youtube.com/watch?v=jFpNTW-id94&t=1s

**Video S2.** Video of AquaGel-K burn played at 3x speed, featuring images of the post-burn substrate. See the video at:
htttps://www.youtube.com/watch?v=DcFD9_gT1Io